\documentclass[10pt,twocolumn,superscriptaddress,showpacs,preprintnumbers,amsmath,amssymb,prb,aps]{revtex4-1}
\usepackage{color}
\usepackage{graphicx,bm}
\usepackage{amssymb,amsmath}
\begin{document}

%%%%%%%%%%%%%%%%%%%%%%%%%%%%%%%%%%%%%%%%%%%%%%%%%%%%%%%%%%%%%%%%%%%%%
\title{Synthesis and characterisation of nanomaterials of the topological crystalline insulator SnTe.}

\author{M. Saghir}
\affiliation{Department of Physics, University of Warwick, Coventry, CV4 7AL, United Kingdom}
\email{M.Saghir@warwick.ac.uk}
\author{M. R. Lees}
\affiliation{Department of Physics, University of Warwick, Coventry, CV4 7AL, United Kingdom}
\author{S. J. York}
\affiliation{Department of Physics, University of Warwick, Coventry, CV4 7AL, United Kingdom}
\author{G. Balakrishnan}
\affiliation {Department of Physics, University of Warwick, Coventry, CV4 7AL, United Kingdom}
\email{G.Balakrishnan@warwick.ac.uk}

%%%%%%%%%%%%%%%%%%%%%%%%%%%%%%%%%%%%%%%%%%%%%%%%%%%%%%%%%%%%%%%%%%%%%
%% The document title should be given as usual. Some journals require
%% a running title from the author: this should be supplied as an
%% optional argument to \title.
%%%%%%%%%%%%%%%%%%%%%%%%%%%%%%%%%%%%%%%%%%%%%%%%%%%%%%%%%%%%%%%%%%%%%
\title{Synthesis and characterisation of nanomaterials of the topological crystalline insulator SnTe.}

%%%%%%%%%%%%%%%%%%%%%%%%%%%%%%%%%%%%%%%%%%%%%%%%%%%%%%%%%%%%%%%%%%%%%
%% Some journals require a list of abbreviations or keywords to be
%% supplied. These should be set up here, and will be printed after
%% the title and author information, if needed.
%%%%%%%%%%%%%%%%%%%%%%%%%%%%%%%%%%%%%%%%%%%%%%%%%%%%%%%%%%%%%%%%%%%%%
%\abbreviations{TI,TCI,ARPES,EDAX,EBSD,SEM,VLS,XRD}
%\keywords{Topological Crystalline Insulator,Vapour-Liquid-Solid, Tin Telluride}

%%%%%%%%%%%%%%%%%%%%%%%%%%%%%%%%%%%%%%%%%%%%%%%%%%%%%%%%%%%%%%%%%%%%%
%% The manuscript does not need to include \maketitle, which is
%% executed automatically.
%%%%%%%%%%%%%%%%%%%%%%%%%%%%%%%%%%%%%%%%%%%%%%%%%%%%%%%%%%%%%%%%%%%%%
%\begin{document}
\begin{abstract}
  A new class of materials, Topological Crystalline Insulators (TCIs) have been shown to possess exotic surface state properties that are protected by mirror symmetry. These surface features can be enhanced if the surface-area-to-volume ratio of the material increases, or the signal arising from the bulk of the material can be suppressed. We report the experimental procedures to obtain high quality crystal boules of the TCI, SnTe, from which nanowires and microcrystals can be produced by the vapour-liquid-solid (VLS) technique. Detailed characterisation measurements of the bulk crystals as well as of the nanowires and microcrystals produced are presented. The nanomaterials produced were found to be stoichiometrically similar to the source material used.  Electron back-scatter diffraction (EBSD) shows that the majority of the nanocrystals grow in the vicinal $\left\{\text{001}\right\}$ direction to the growth normal. The growth conditions to produce the different nanostructures of SnTe have been optimised. 
\end{abstract}

\maketitle
\section{Introduction}

The investigation of topological insulators (TIs) in recent years has led to the discovery of 2D and 3D materials exhibiting exotic surface states (such as HgTe, Bi$_2$Se$_3$ and Bi$_2$Te$_3$).\cite{Hasan2010a,Zhang2000} Topological crystalline insulators (TCIs) are a new subset of materials first predicted through theory,\cite{Tanaka2012,Fu2011} and later observed in experiments on bulk crystals of the IV-VI semi-metal SnTe showing band degeneracy.\cite{Tanaka2012} SnTe is a narrow-gap IV-VI semiconductor with a rock salt cubic structure (lattice constant, \emph{a} = 0.63 nm). The band degeneracy observed in TCIs are protected by rotational and mirror symmetry, in place of the role played by time-reversal symmetry in TIs.\cite{Hsieh2008,Hsieh2009,Xia2009} Angle-resolved photoemission spectroscopy (ARPES) is one measurement technique that can be used to reveal the exotic surface states of these materials. Measurements to date have been performed only on the surfaces of cleaved bulk crystals or thin films.\cite{Kong2010a} ARPES performed on SnTe has been reported to show Dirac cone surface states.\cite{Hsieh2012}

It has been reported that TI materials in nanoform allow the observation of enhanced TI features due to an increased surface-area-to-volume ratio.\cite{Cha2010} As materials are transformed into their nano counterparts, the surface states have a greater contribution to the sample properties and conversely, bulk properties are suppressed.\cite{Kong2010a} To date, many successful growths of nanomaterials have been performed for both binary and ternary compounds of the Bi$_2$Se$_3$/Te$_3$ family. High quality nanomaterials can be obtained with various stoichiometries using both wet and dry synthesis methods.\cite{Kong2010b,Kong2010a,Wang2013,Li2011} Some examples of synthesis methods include van der Waals epitaxy of Bi$_2$Te$_2$Se on hexagonal boron nitride sheets or electro deposition of Bi$_2$Te$_2$Se/Te to form multi-arrays of nanowires.\cite{Gehring2012,Li2011} Another common method for nanowire growth is similar to that for the fabrication of ZnO nanowires using vapour-liquid-solid (VLS) technique.\cite{Yang2002} The VLS growth technique is known to be an effective growth method for high quality crystalline nanomaterials with high yields.\cite{Lee2008,Lieber2011,Yang2005,Gao2002,Medlin2010} The VLS method has also been used to demonstrate the growth of Bi$_2$Se$_3$ and Bi$_2$Te$_3$ nanomaterials.\cite{Wei2010,Kong2010b,Cha2010} SnTe has been previously reported to have been produced in nanoform using methods such as chemical reduction, hydrothermal synthesis as well as chemical synthesis.\cite{Leontyev2012,Salavati-Niasari2010,Kovalenko2007} We have also become aware of the recent publications by Li \emph{et al} and Safdar \emph{et al} for the growth of nanostructures of SnTe using the VLS method.\cite{Li2013,Safdar2013}

Here we report experimental evidence for the growth of SnTe nanostructures using a Au-catalysed VLS growth technique starting with large single crystal boules of SnTe. The optimum growth conditions are presented to obtain nanowires and microcrystals of SnTe. Characterisation performed on the materials grown include powder x-ray diffraction (XRD), x-ray Laue diffraction, scanning electron microscopy (SEM), energy dispersive x-ray analysis (EDAX) and EBSD. 

\section{Results and discussion}

We begin by detailing the procedure for the growth of bulk crystals of SnTe, followed by those used for the nanowires and microcrystals of SnTe. By following the procedure reported by Tanaka \emph{et al}, crystal boules of SnTe were produced using a modified Bridgman.\cite{Tanaka2012} Several samples of SnTe were synthesised starting from high purity elements (99.99 \% pure) which were mixed in varying ratios (Sn:Te = 51:49, 50:50 and 49:50) and then sealed in evacuated quartz tubes. The quartz tubes were placed vertically into a box furnace and the contents melted at  900 $^\circ$C. The quartz tubes remained at this elevated temperature for two days. The tubes were then slow cooled (2 $^\circ$C/h) to 770 $^\circ$C and then rapidly cooled down to room temperature. To check the phase purity, a small section of the boule was finely ground and powder x-ray diffraction patterns were obtained using a Panalytical X'Pert Pro system with monochromatic CuK$\alpha_1$ radiation. X-ray Laue diffraction was performed on the crystal boules and across the faces of cleaved sections of the boule to check the quality of the crystals produced. Finally, the stoichiometry was determined using an EDAX EBSD system installed alongside a Zeiss SUPRA 55-VP scanning electron microscope.

\begin{figure}
\centering
\includegraphics[width=0.5\textwidth]{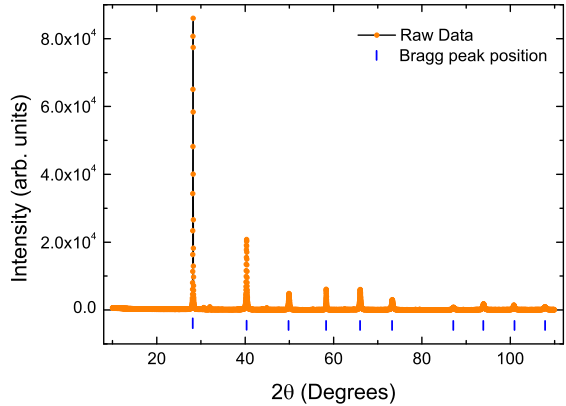}
\caption{Powder x-ray diffraction pattern of a crushed sample of the as-grown SnTe crystal (composition 51:49). The data obtained (orange) and the expected Bragg peak positions (blue) are shown.}
\label{fig:XRD}
\end{figure}

Upon removal of the bulk crystal boules from the sealed quartz tubes, a visual inspection of the crystal boules revealed shiny metallic surfaces. The crystal boules cleaved easily revealing flat mirror surfaces. A small section of the crystal boule was isolated and ground into a fine powder for characterisation using powder XRD. The data obtained for the 51:49 composition shows that the SnTe was single phase with no impurities as seen in figure \ref{fig:XRD}.
\begin{figure}
\centering
\includegraphics[width=0.2\textwidth]{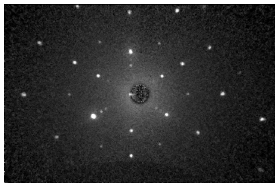}
\caption{Typical Laue pattern of the as-grown crystal boule. The incident x-ray beam is aligned with the [111] crystallographic direction. The Laue patterns were consistent across the surface of the boule.}
\label{fig:Laue}
\end{figure}
It was found that the majority of crystals cleaved along the $\left\{\text{100}\right\}$ and $\left\{\text{111}\right\}$ planes. Compositional analysis using EDAX was performed on several small pieces of crystal obtained from different regions of the boule. The data revealed that the crystals had formed stoichiometrically as can be seen in Table \ref{tab:EDAX}.
Laue diffraction (figure \ref{fig:Laue}) was performed on several freshly cleaved surfaces which revealed well defined sharp diffraction spots.
\begin{table}
\caption{Representative atomic compositions of the bulk SnTe crystal boules, nanowires and microcrystals obtained using EDAX analysis.} % title of Table
\centering % used for centering table
\begin{tabular}{c c c c} % centered columns (4 columns)
\hline \\[-1.5ex]%inserts double horizontal lines
 & Sn & Te & Total \\ [0.5ex] % inserts table
%heading
% inserts single horizontal line
\hline \\[-1.5ex]
SnTe Boule & 51.5(5) & 48.5(5) & 100 \\  % inserting body of the table
Nanowires & 50.5(5) & 49.5(5) & 100 \\ % inserting body of the table
Microcrystals & 50.3(5) & 49.7(5) & 100 \\  [1ex] % [1ex] adds vertical space
\hline % inserts single horizontal line
 % inserts single horizontal line
\end{tabular} 
\label{tab:EDAX} % is used to refer this table in the text
\end{table}

\begin{figure*}
\centering
\includegraphics[width=0.8\textwidth]{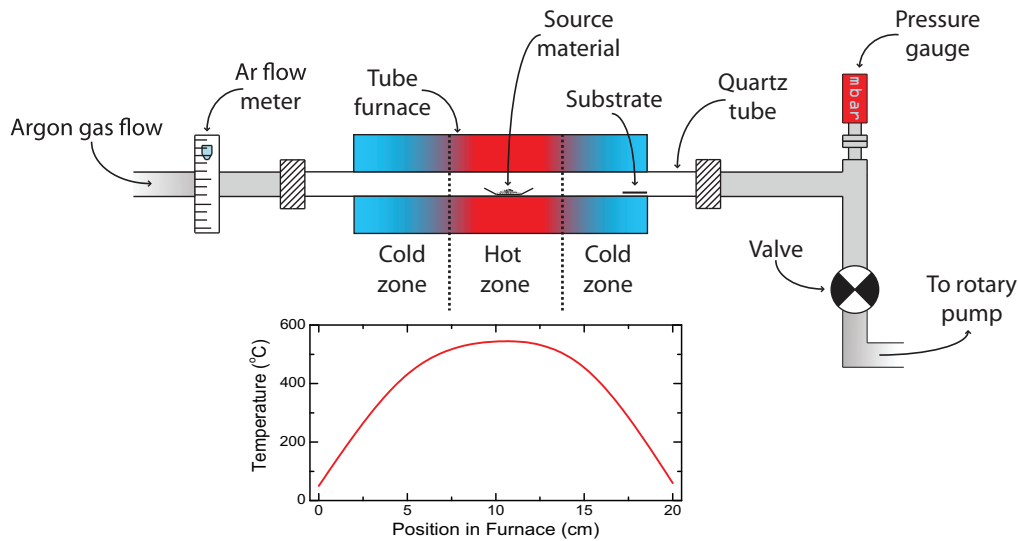}
\caption{Schematic of the quartz tube furnace showing the positions of the source material and the substrate. Bottom panel: The measured temperature profile along the length of the furnace for a set temperature of 540 $^\circ$C at the hot zone.}
\label{fig:Schematic}
\end{figure*}

Two different growth procedures for the nanomaterials were attempted resulting in variable morphologies of SnTe. These included nanowires, stacks and microcrystals, depending on the position and therefore the temperature of the substrate downstream in the tube furnace used for the growth. For the growth of nanomaterials, a schematic of the quartz tube furnace used and the experimental arrangement used is shown in figure \ref{fig:Schematic}.
Silicon substrates were prepared using two methods depending on the SnTe morphologies desired. The first was found to be appropriate for the formation of nanowires. This involves using Au nanoparticles as a precursor for nanowire growth. The second method was found to promote the growth of microcrystals and required using a smooth Au layer deposited onto the substrate used. Over 40 experiments were conducted, varying a number of parameters, such as substrate position, source material position, temperatures of the hot and cold zones, Ar gas flow rates used for the growth and duration of growth. The methods reported below explain the processes used to obtain nanomaterials of high yields and the optimum conditions for the growth and crystallinity.
For both methods, a small portion of the SnTe boule (0.05 g) was ground into a fine powder and placed in an alumina-silicate boat to act as the source material for the growth. In most of the studies reported here, the boule with the nominal starting composition 51:49 was used. The silicon substrates used were 50 mm $\times$ 5 mm strips of silicon wafer, cleaved and cleaned using a solvent mixture of 50:50 isopropan-2-ol and acetone. The silicon substrate was blow dried using nitrogen gas.

For the first method, the silicon substrate was dipped into and immediately removed from a sodium citrate solution containing 20 nm gold nanoparticles (Sigma Aldrich). With the substrate placed horizontally, the solution was allowed to evaporate in air at room temperature for approximately 30 minutes. As the substrates dried, evenly spaced gold nanoparticles were observed across the substrate surface using an SEM. The alumina silicate boat containing the SnTe powder was placed in the centre `hot-zone' of the quartz tube furnace. The prepared silicon substrate was placed downstream in the furnace `cold-zone'. The quartz tube was evacuated and flushed several times with high purity argon gas (99.997 \%) to ensure that there was an inert atmosphere in the chamber prior to growth. A steady flow of argon gas of 35 sccm was then established to act as the carrier gas and to reduce oxidation and contamination during the growth process. The furnace temperature was ramped up rapidly from room temperature (in 10 mins) to reach a target temperature of 540 $^\circ$C at the hot zone, while the temperature at the cold zone where the substrate was placed was maintained $\approx$ 300 $^\circ$C. The hot zone of the furnace was maintained at 540 $^\circ$C for a period of 120 mins, while the growth was allowed to take place and after this period, it was cooled to room temperature. The furnace was calibrated using an external temperature probe that was positioned at various fixed locations along the length of the tube, for a fixed temperature (set point) at the hot zone. From this, a temperature profile over the entire length of the tube inside the furnace was obtained.

For the second procedure, a thermal evaporator was used to deposit a 50 to 100 nm layer of gold onto the silicon substrate to act as a catalyst for growth. The substrate and source material were then placed inside the quartz tube furnace as shown in figure \ref{fig:Schematic} and the same growth conditions were employed as described for the first method.
Samples obtained were characterised using SEM, EDAX and EBSD.

After the VLS growth process was completed, a metallic coating on the inside of the quartz furnace tube was formed downstream and the substrate appeared to show metallic crystalline features on the surface when viewed with the naked eye. The substrates were then inspected using SEM. We found that the two different procedures adopted, gave different results. It was found that for the first method, long smooth single crystal nanowires were observed and for the second procedure, where gold was sputtered on the substrate, the structures formed were microcrystals.
%\begin{flushleft}
%\text{C.2.1 SnTe Nanomaterials}
%\end{flushleft}
\begin{figure*}[t]
\centering
\includegraphics[width=1\textwidth]{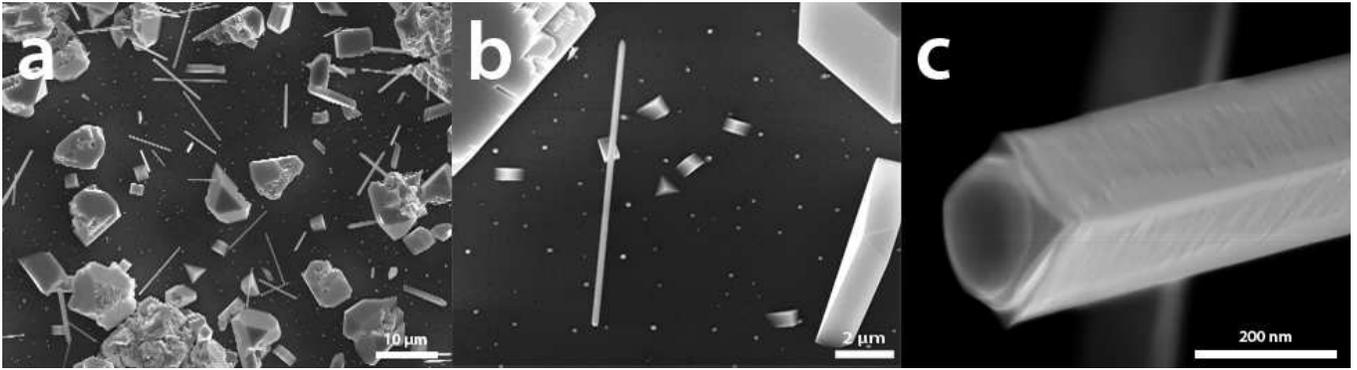}
\caption{(a) SEM image showing a high density of SnTe nanowires amongst microcrystals. (b) SEM image showing a typical nanowire seen protruding from the sample surface surrounded by evenly spaced Au-nanoparticles. (c) High resolution SEM image of the tip of a nanowire showing the smooth nature of the growth. The end of the nanowire can be seen to contain the alloyed Au-nanoparticle which travels along the growth direction.}
\label{fig:SEM}
\end{figure*}

We first describe the results obtained on the nanowires. Representative images of the substrate surface for the first procedure are shown in figure \ref{fig:SEM} along with a high resolution image of an isolated nanowire. Nanowires were observed to grow in random orientations protruding outwards from the sample surface. Typically, the nanowires grown are 200 nm wide and 20 to 50 $\mu$m long. EDAX analysis of the nanowire shows that within experimental error the composition is stoichiometrically similar to the source material used. The composition of the nanowires determined using EDAX is presented in Table \ref{tab:EDAX}. From the high resolution SEM image of a single isolated nanowire in figure \ref{fig:SEM}, it can be seen that the Au-nanoparticle terminates the growth of the nanowire suggesting that it promotes growth and that growth occurs below the Au-nanoparticle.

This is similar to the recent reports by Li \emph{et al}.\cite{Li2013} for this material. The diameter of the nanowires can also be seen to be constrained to the width of the Au-nanoparticle alloy formed, as all nanowires observed were $\approx$ 200 nm $\times$ 200 nm in cross-section. It was reported in ref. 25 that the nanowires they observed were wider/thicker than the alloy particle, which we see no evidence of. Nanowire formation was best observed in the center of the substrate with microcrystals forming at the warmer end of the substrate (i.e. closer to furnace centre/hot zone).

As a control, a growth was performed without the use of Au-nanoparticles and no nanowires were observed, which suggests the importance of using Au-nanoparticles to catalyse the growth of nanowires. In the growth procedures described by Li \emph{et al}., the temperature of source material as well as the substrate was much higher than our reported growth temperatures. Their growth periods were also much shorter (30 mins). Our experiments show that when varying both the growth temperature as well as the duration over a very wide range, no evidence of nanowire or microcrystal formation is observed at these high temperatures and for shorter durations.

\begin{figure}[!htbp]
\centering
\includegraphics[width=0.35\textwidth]{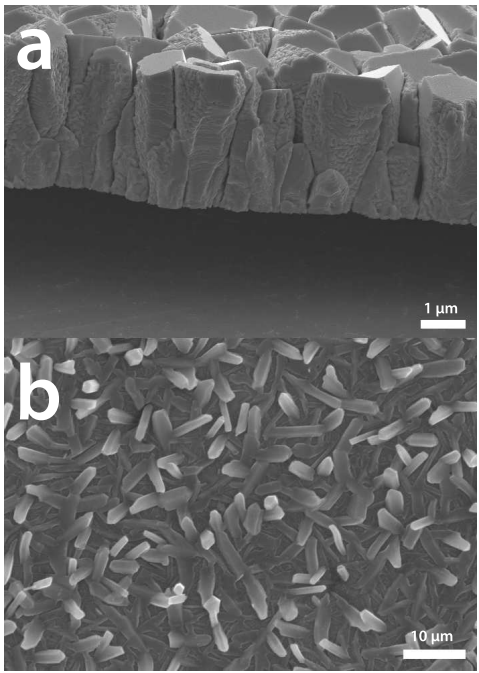}
\caption{(a) Representative SEM image of the growth of SnTe for periods longer than 120 minutes. The thickness of the SnTe film grown is $\sim$ 15 $\mu$m (b) Representative SEM image of the growth of SnTe for temperatures greater than the optimum temperature of 540 $^\circ$C. Much greater nucleation is observed coupled with a greater growth rate giving rise to thicker structures resembling nanowires.}
\label{fig:SnTeVarying}
\end{figure}

We find that our optimum growth temperature is 540 $\pm$ 20 $^\circ$C and duration is $\sim$120 mins. For periods longer than 120 mins, we observe a more complete coverage of SnTe. A layer 15 $\mu$m thick across the silicon substrate surface is formed and at higher temperatures we observe more nucleation (figure \ref{fig:SnTeVarying}). For periods less than 120 mins, very little deposition was seen, as was the case at lower temperatures.

\begin{figure}[!htbp]
\centering
\includegraphics[width=0.4\textwidth]{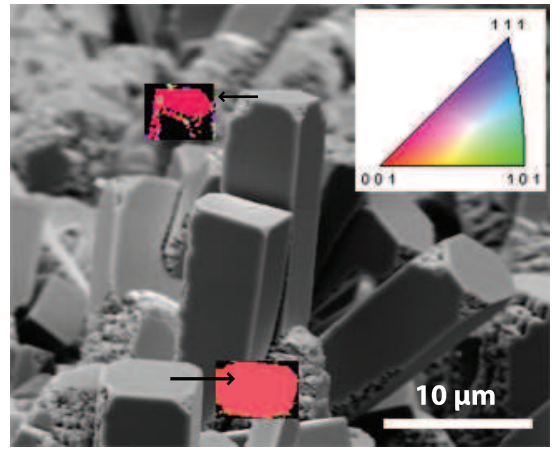}
\caption{SEM image of SnTe microcrystals. The two red insets placed next to the corresponding microstructures show the EBSD pattern of the growth orientation in the direction normal to the growth axis. This is typically found to be a vincinal $\left\{\text{001}\right\}$ orientation for the majority of crystals but some show a $\left\{\text{111}\right\}$ growth plane.}
\label{fig:EBSD}
\end{figure}

The second experimental procedure described above was conducive to the growth of microcrystals. The morphology of SnTe structures found in this procedure were microcrystal rods and stacks. A typical SEM image obtained is shown in figure \ref{fig:EBSD}.
Perfectly formed cubic crystal structures of SnTe can be seen. Using EDAX, these were determined to be stoichiometric in composition to the source material. Microcrystal rods were found to form in clusters on the edges of the substrate, protruding out of the surface. They also formed as single stacks on the surface as flat planes parallel to the substrate surface. Using EBSD, it was found that all the structures grew predominantly in the vicinal $\left\{\text{001}\right\}$ growth orientation as can be seen in figure \ref{fig:EBSD}. This was also observed by Li \emph{et al}. where the growth direction of their crystals is along the $\left\langle 001\right\rangle$ zone axis. It was also found that the Au-layer, sputtered onto the substrate surface, played an important role in promoting growth as no ordered SnTe structures were observed when the same growth procedure was carried out without the Au layer.

In summary, we have determined the optimum growth conditions for the formation of nanowires and microcrystals of SnTe. Both these structures were produced using the VLS growth method. We have determined the growth orientation of SnTe microcrystals using EBSD and found this to be vicinal $\left\{\text{001}\right\}$ to the direction normal. We find that a Au precursor is essential for the growth of both the nanowires as well as the microcrystals. The growth results are dependent on the temperature of the source material, the substrate temperature which corresponds to the position of the substrate in the cold zone and the growth duration. This demonstration for the growth of topological insulators and their derivatives in micro or nanoform allows a pathway to be established to potentially exploit and enhance the observable topological surface state features by reducing the bulk contribution.

\begin{acknowledgements}

This work was supported by the EPSRC, UK (EP/I007210/1). Some of the equipment used in this research was obtained through the Science City Advanced Materials Project, Creating and Characterizing Next Generation Advanced Materials Project, with support from Advantage West Midlands (AWM), and was partially funded by the European Regional Development Fund (ERDF). The authors thank M. Ciomaga Hatnean for the x-ray analysis and T. E. Orton for valuable technical support. We thank Dr N. R. Wilson for useful discussions.

\end{acknowledgements}

\bibliography{CGD3}

\end{document}